# Spatial Symmetry of Superconducting Gap in YBa$_2$Cu$_3$O$_{7-\delta}$ Obtained from Femtosecond Spectroscopy

C. W. Luo[1], M. H. Chen[1], S. P. Chen[1], K. H. Wu[1], J. Y. Juang[1],
J.-Y. Lin[2], T. M. Uen[1] and Y. S. Gou[1]

[1] *Department of Electrophysics, National Chiao Tung University, Hsinchu, Taiwan, R.O.C.*
[2] *Institute of physics, National Chiao Tung University, Hsinchu, Taiwan, R.O.C.*

The polarized femtosecond spectroscopies obtained from well characterized (100) and (110) YBa$_2$Cu$_3$O$_{7-\delta}$ thin films are reported. This bulk-sensitive spectroscopy, combining with the well-textured samples, serves as an effective probe to quasiparticle relaxation dynamics in different crystalline orientations. The significant anisotropy in both the magnitude of the photoinduced transient reflectivity change and the characteristic relaxation time indicates that the nature of the relaxation channel is intrinsically different in various axes and planes. By the orientation-dependent analysis, *d*-wave symmetry of the bulk-superconducting gap in cuprate superconductors emerges naturally.

The crucial and yet controversial issues involved in the superconducting gap symmetry of the cuprate superconductors remain unsettled. Numbers of theories and experiments have been proposed and deployed to track down the nature of the order parameter symmetry of the cuprate superconductors. In general, the obtained results can be roughly classified into two sects[1]. Majority of experimental results obtained by surface-sensitive experiments[2-6], all point to the widely-accepted *d*-wave symmetry scenario. Nonetheless, recently some bulk-sensitive experiments[7-10], e.g. the investigation of quasiparticle dynamics by the ultrafast time-resolved experiments, have revealed some disputable features of *s*-wave symmetry or *s-d* mixed characters. In particular, very recently, Kabanov *et al.*[7] compared the calculations on the temperature dependence of the photo-induced transmission amplitude below $T_c$ and claimed that the results were more consistent with an isotropic gap in YBa$_2$Cu$_3$O$_{7-\delta}$ (YBCO). This apparently has revived the extensive interest on this matter. The fact that, in addition to its bulk-sensitive characteristic, there exist intimate correlations between the superconducting gap opening and the amplitude as well as relaxation time of the transient reflectivity change (ΔR/R) has made the femtosecond pump-probe technique one of the most powerful tools in studying the ultrafast carrier dynamics relevant to high-$T_c$ superconductivities[11-19]. It is, thus, desirable to re-examine this issue with the improved femtosecond laser and more carefully characterized samples.

In this study, based on the general consensus that *the amplitude and relaxation time of the transient reflectivity in picosecond scale below $T_c$ are directly associated with the opening of the superconducting gap*, we have measured the orientation-dependent ΔR/R in the *bc-*, *ab*-diagonal-*c*, and *ab*-planes of YBCO by polarized femtosecond time-resolved spectroscopy. The independent polarization of the pump and probe beams and the well-oriented films together provide detailed information that is unprecedented in the literature. The significant anisotropy in both the magnitude and the characteristic relaxation time of ΔR/R indicates that the nature of relaxation channel might be intrinsically anisotropic in different planes and *even along different orientations in the CuO$_2$ (ab) planes*.

The well characterized (100)- and (110)-oriented YBCO thin films were used in this study. Both samples were prepared by pulsed laser deposition. In order to obtain the desired epitaxial orientations, the substrate, deposition temperature, and even the oxygen partial pressure were carefully monitored. For (110)-oriented films, SrTiO$_3$ (110) substrates were used with a buffer layer of (110) PrBa$_2$Cu$_3$O$_{7-\delta}$ deposited at lower temperatures. This process not only successfully suppressed the growth (103) YBCO but also improved the $T_c$ significantly[20]. For preparing the highly in-plane aligned (100) YBCO films, a novel process was deployed[21].

The percentage of the in-plane alignment for both (100) and (110) films was larger than 97% as determined by the x-ray Φ scanning. In order to distinguish the directions of the axes in the thin film plane, the polarized x-ray absorption near-edge spectroscopy (XANES) of the O *K*-edge was carried out and successfully identified the *a*-, *b*-, and *c*-axes of YBCO thin films[21, 22]. The orientations of both the (100) and (110) films were further checked by the electric resistivity measurements[23] and were consistent with those obtained by XANES. The zero resistance transition temperatures ($T_{c0}$) are 89.7 K and 88.2 K for (100) and (110) YBCO films, respectively.

Samples were mounted on the cold finger of a Janis flow-through cryostat and the temperature was monitored by the thermometer buried in the finger below samples. The femtosecond time-resolved spectroscopy was carried out by the usual pump-probe scheme. The pump-probe measurements utilized a mode-locked Ti: sapphire laser which produced a 75 MHz train of 20 fs pulses with a central wavelength of 800 nm. The ratio between the average power of the pump and probe beams was set at 40:1. The typical energy of the pump pulses was ~0.5 nJ and modulated at 87 KHz with an AO modulator. The small reflecting signals were detected by a lock-in amplifier. In practice, the pump pulse excites electron-hole pairs that relax to the states in the vicinity of the Fermi energy ($E_F$) by either electron-electron or electron-phonon scattering. This process occurs within a time scale of sub-picosecond[7, 24]. The presence of a gap near $E_F$ leads to the carrier accumulation in quasiparticle states above the gap. This, in turn, gives rise to a transient change in reflectance (ΔR/R(*t*)) to be detected by a second laser (probe) pulse as a function of a time delay between the pump and probe pulses. The amplitude and characteristic relaxation time of the measured ΔR/R(*t*) thus give important information of the number of the quasiparticles ($N_{QP}$) accumulated and the amplitude of the gap, according to |ΔR/R| ∝ $N_{QP}$ ∝ 2Δ (superconducting gap)[7, 12, 13]. To achieve the orientation-resolved femtosecond spectroscopy,

the polarization-dependent pump-probe experimental setup is shown in Fig. 1. The ΔR/R ($t$, $\Phi_1$, $\Phi_2$, $\theta$) curves along various directions on the surface of the sample can be obtained by rotating the polarization (electric field, **E**) of pulses ($\Phi_1$, $\Phi_2$) at nearly normal incidence, $\theta \sim 0$ ° (actually, $\theta_{pump}$ = 4 ° and $\theta_{probe}$ = 1.5 °). With the polarization of the pulses perpendicular to the $c$-axis of (100) or (110) films, one is able to measure the responses (ΔR/R ($t$, 0, 0, $\theta$)) along various directions in the $ab$-plane by changing the angle $\theta$. It is repeatedly checked and confirmed that at any particular probing orientation the prominent characteristics of ΔR/R only depend on the direction of **E**$_{probe}$ while the independently manipulated **E**$_{pump}$ only affects the excitation efficiency, i.e. the amplitude of ΔR/R[24, 25]. Thus, in order to obtain the largest effect, we adopt the **E**$_{pump}$//**E**$_{probe}$ configuration to investigate the anisotropic properties of YBCO in this study.

Fig. 2(a) shows the typical ΔR/R in the superconducting state by setting the polarizations of the pump and probe pulses parallel to the $b$-, $c$-axes, and $ab$-diagonal directions. The ΔR/R along the $b$-axis is markedly different from the other two, not only in the amplitude, but also in the characteristic of relaxation. Two relaxation processes (fast component, $\tau_1$ and slow component, $\tau_2$) can be clearly observed along the $b$-axis, indicating the occurrence of a gap near $E_F$[26, 27]. On the contrary, the slower relaxation process is absent in both the $c$-axis and $ab$-diagonal. Furthermore, the amplitude difference of ΔR/R is about 5-10 times larger in $b$ than in the other two directions. These results clearly demonstrate that cuprate superconductors are indeed very anisotropic and the superconducting gap appears to emerge only in certain crystalline orientation, namely [010] or $b$-axis.

By scanning the polarization of both the pump and probe pulses ($\Phi_1$ = $\Phi_2$ = 0 ° to 90 ° and setting $\theta$ = 0 °), a complete picture of the spatial symmetry of superconducting gap can be mapped in both $bc$-plane and $ab$-diagonal-$c$ plane, as illustrated in Fig. 3. In the normal state, no significant change of ΔR/R was observed along various directions in the $bc$-plane (see Fig. 3(a) and the insets, $T$ = 120 K). In the superconducting state, however, a large number of the photoexcited quasiparticles (~ 1.66×10$^{11}$ /per pulse, based on the assumption that each photon creates 2×$h\nu$/2Δ quasiparticles with 2Δ ~ 40 meV being the order of superconducting gap) are generated by the pump pulses along the $b$-axis and followed a slow relaxation process due to the bottleneck set by the opening of superconducting gap[28]. Off the $b$-axis, the number of the accumulated quasiparticles through the photoexcitation process dramatically drops (ΔR/R decreases) and $\tau_2$ significantly increases from ~4.7 ps to ~6.5 ps (see Fig. 3(a) and the insets, $T$ = 60 K). This indicates that the recombination of quasiparticles becomes less and less efficient due to the shrinking of the superconducting gap[7, 26]. Along the $c$-axis, ΔR/R becomes very small and $\tau_2$ disappears completely, indicative of the absence of the superconducting gap. This characteristic is rather different from that in the $ab$-diagonal-$c$ plane. In that case, even in the superconducting state ($T$ = 60 K), no significant change of ΔR/R was observed over the entire scan as shown in Fig. 3(b) and the insets. This is also very similar to that in the normal state ($T$ = 100 K, right panel in Fig. 3(b)). These results are consistent with our previous conjecture that the superconducting gap exists only in the vicinity along the *b*-axis and is nearly absent in the *c*-axis and *ab*-diagonal directions.

The spectroscopy in the *ab*-plane is of particular interest. By using *(110) and (100) YBCO thin films* and rotating the angle of sample ($\theta_s$) and setting $\Phi_1 = \Phi_2 = 0°$ at 60 K, the characteristics of the superconducting gap along various directions in the *ab*-plane are displayed in Fig. 2(b). For abnormal incidence cases, we have calibrated the actual angles ($\theta$) of the optical polarization inside YBCO films by Snell's law with $n_{YBCO} = 1.64$. It is evident that, along *ab*-diagonal ($\theta \sim 45°$), the diminishing $\Delta R/R$ amplitude suggests a minimal number of the accumulated quasiparticles in the metastable states. While for other directions, the amplitude of $\Delta R/R$ indicates that the quasiparticles are easier to accumulate. As mentioned before, the $\Delta R/R$ response primarily depends on the polarization of the probe beam, and the symmetry of the superconducting gap can thus be inferred by this type of measurements. This is even better illustrated by the angular dependence of the amplitude and relaxation time of $\Delta R/R$ shown in Fig. 4. It is interesting to note that, in Fig. 4(a), the amplitude of $\Delta R/R$ drops drastically at $\theta = 45°$ and the data are qualitatively consistent with the $d_{x^2-y^2}$-wave symmetry depicted by the dashed pattern shown in Fig. 4. Similarly, the diverge of $\tau_2$ (open circles in Fig. 4(b)) when approaching *ab*-diagonal is indicative of a shrinking superconducting gap leading to the inefficient quasiparticle recombination[7, 26]. Moreover, the abrupt disappearance at the *ab*-diagonal strongly suggests the existence of gapless nodes along the *ab*-diagonal. These results lend strong support to the conjecture that the intrinsic nature of the bulk-superconducting gap is indeed dominated by *d*-wave symmetry.

Finally, we would like to note that it is virtually impossible to resolve the spatial symmetry of the superconducting gap in the *ab*-plane by using the (001) YBCO thin films, since in this case the *a*- and *b*-axes are randomly orientated in the *ab*-plane. The results consist of contribution from both *a*- and *b*-axes directions and there is no way of getting correct temperature-dependent $\Delta R/R$ *along the ab-diagonal* with (001) YBCO films[29, 30]. We believe that the apparent discrepancy between the present results and those obtained with the same femtosecond spectroscopy in (001) YBCO films by Kabanov *et al.*[7] must have arisen from the above factors.

In summary, by using the (100)- and (110)-oriented YBCO thin films, the spatial anisotropy of photoinduced quasiparticle relaxation dynamics in several major crystalline planes can be clearly revealed by the polarized femtosecond spectroscopy. Except in the *ab*-diagonal-*c* plane, spectra in both the *bc*- and *ab*-planes show strong anisotropies of $\Delta R/R$, in contrast to those in the *ab*-diagonal-*c* plane. The results strongly support that the *d*-wave symmetry in the superconducting order parameter of cuprate superconductors not only appears on the surface-sensitive experiments, but also robustly prevails in the bulk. Therefore, it seems to dismiss the possibility that the superconducting wavefunction may vary as a function of the distance from the surface as recently suggested by Müller[1].

This work was supported by the National Science Council of Taiwan, R.O.C. under grant: NSC90-2112-M009-036.

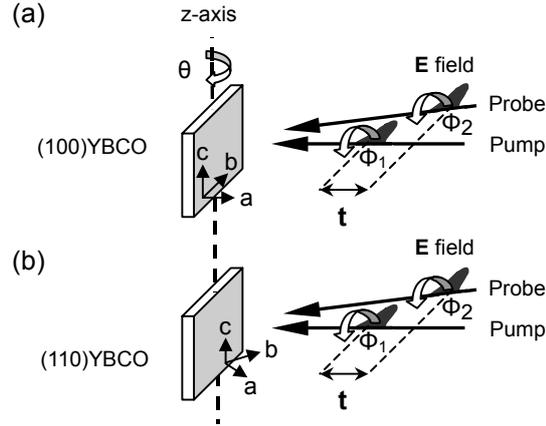

Fig.1. The polarized pump-probe experimental setup. $\Phi_1$ ($\Phi_2$) is the angle between the (a) *b*-axis; (b) *a-b* plane and the pump (probe) pulses. $\theta$: the angle between the surface of samples and the polarization of the pump (or probe) pulses. *t*: the time delay between the pump and probe pulses.

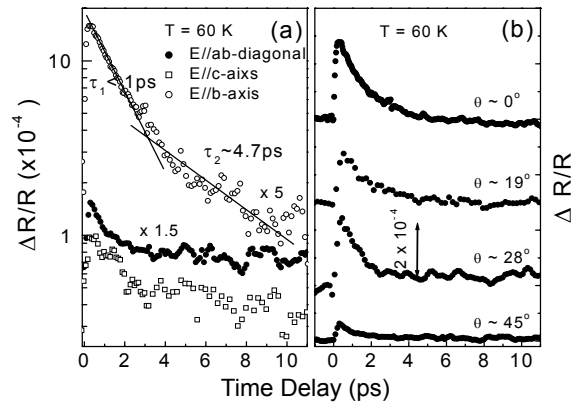

Fig. 2. (a) ΔR/R measured at 60 K along various crystalline orientations. The spectra of *E*//*c* and *b* were measured using (100) YBCO films. Those of *E*//*ab*-diagonal were measured using (110) YBCO films. (b) ΔR/R along specific crystalline directions in the *ab*-plane at 60 K. Results of $\theta \sim 45°$ were measured on (110) YBCO films with nearly normal incidence. Data of other angles were measured on (100) YBCO films.

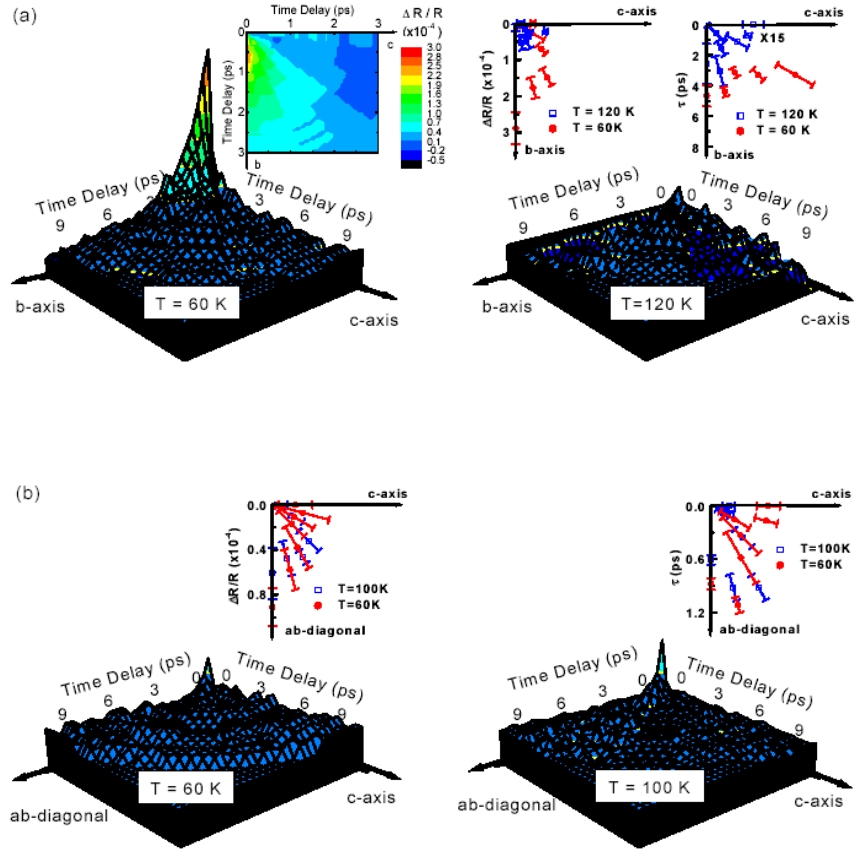

Fig. 3. ΔR/R along various directions is shown for (a) the *bc*-plane of (100) YBCO films at 60 and 120K; (b) the *ab*-diagonal-*c* plane of (110) YBCO films at 60 and 100 K, by rotating the polarization of laser pulses. The insets are the amplitude and relaxation time of the corresponding 3D ones. Note: for the *bc*-plane at 60K, only the slow relaxation time ($\tau_2$) is shown.

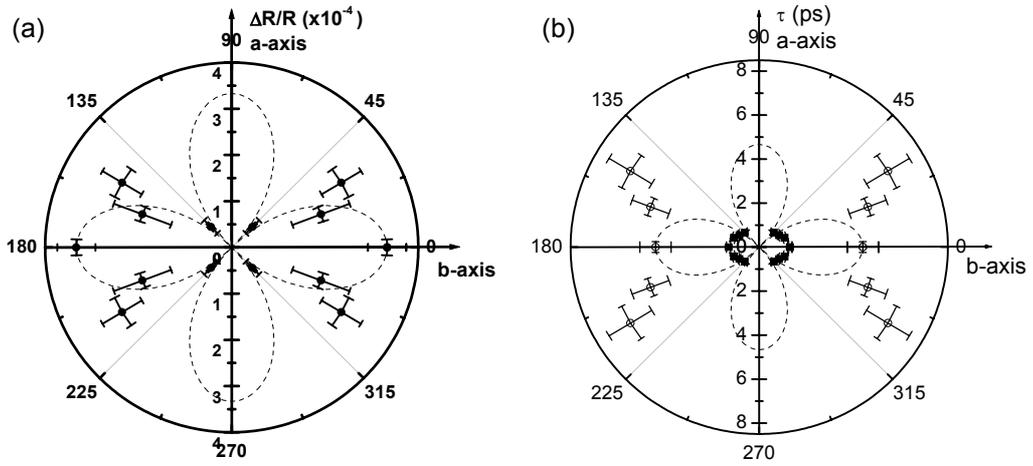

Fig. 4. The angular dependence of (a) the amplitude and (b) the relaxation time, $\tau_1$ (fast component, solid squares) and $\tau_2$ (slow component, open circles), of ΔR/R in the *ab*-plane at 60 K. The dash lines are according to the $d_{x^2-y^2}$-symmetry and normalized to the (a) amplitude and (b) $\tau_2$ of ΔR/R along the *b*-axis, respectively.